%% file: main.tex
\documentclass[11pt]{article}

\usepackage[a4paper,margin=2.5cm]{geometry}
\usepackage{fontspec}
\usepackage{xeCJK}
\setCJKsansfont[BoldFont=HaranoAjiGothic-Bold.otf]{HaranoAjiGothic-Regular.otf}
\setCJKmonofont{HaranoAjiGothic-Regular.otf}

\usepackage{booktabs}
\usepackage{array}
\usepackage{longtable}
\usepackage{pdflscape}
\usepackage{graphicx}
\usepackage{enumitem}
\usepackage{xcolor}
\usepackage{listings}
\usepackage{tikz}
\usetikzlibrary{positioning,shapes.geometric,arrows.meta}
\usepackage[hidelinks]{hyperref}

\newcolumntype{P}[1]{>{\raggedright\arraybackslash}p{#1}}

\newsavebox{\tabbox}
\newenvironment{tablenote}[1][\wd\tabbox]
  {\smallskip\par\noindent\begin{minipage}{#1}%
     \setlength{\parindent}{0pt}%
     \leftskip=0pt\rightskip=0pt\parfillskip=0pt plus 1fil\relax
     \footnotesize}
  {\par\end{minipage}}

\title{\texttt{tse\_tick}: A Python Library for Parsing and Querying\\
Nikkei NEEDS Tick Data from the Tokyo Stock Exchange}

\author{
  Kazumi Li$^{1}$ \quad Masataka Hayashi$^{2}$ \quad Teruo Nakatsuma$^{2}$ \quad Peter Romero$^{3}$\\[8pt]
  {\small $^{1}$Graduate School of Economics, Keio University}\\
  {\small $^{2}$Faculty of Economics, Keio University}\\
  {\small $^{3}$Psychometrics Centre, University of Cambridge}\\[4pt]
  {\small Corresponding author: \texttt{kaiwenli@keio.jp}}
}
\date{}

\begin{document}
\maketitle

\begin{abstract}
\noindent
Tick-level trade-and-quote data for the Tokyo Stock Exchange is distributed
through the Nikkei NEEDS service as thousands of zipped CSV archives spanning
four data types with era-dependent schemas and Japanese-language layouts. We
present \texttt{tse\_tick}, an open-source Python library that converts these
raw archives into clean, typed Polars DataFrames and a Hive-partitioned
Parquet store queryable through DuckDB. The library offers two access paths
sharing one parse-and-clean core: a one-shot reader that returns a ticker- and
time-filtered DataFrame directly from raw ZIP files, and a two-stage
ingest-then-query pipeline with resume-safe, memory-aware parallel ingestion,
part-pruning, and a materialized intraday time key for row-group pruning. The
engineering, more than the parsing, is what the library contributes: ingestion
runs in per-date atomic units whose completion is recorded by coverage markers
rather than inferred from file existence, writes stream in bounded morsels so
that peak memory is independent of trading-day size (24.5 GB to 2.4 GB on the
worst measured day), a RAM-aware process pool sizes itself to available
memory, and part-pruning opens only the archive parts a ticker can occupy.
Full English and Japanese column definitions ship for all four types, and a
translation layer maps yfinance, Polygon, and ccxt names onto their
\texttt{tse\_tick} equivalents. In benchmarks on a commodity 16-thread
workstation, parsing a representative 4.8-million-row archive part, one of a
trading day's nine parts, is 59.8$\times$ faster than the original pandas
prototype (34.3$\times$ against an engine-matched pandas baseline), and a
single-ticker time-window query from the store completes roughly 410$\times$
faster than a pandas scan of the equivalent CSV. \texttt{tse\_tick} is
available on PyPI (\texttt{pip install tse-tick}) under the MIT license.
\end{abstract}

\section{Introduction}
\label{sec:introduction}

Market-microstructure research depends on tick-level data, and for the Tokyo
Stock Exchange (TSE) the standard institutional source is the Nikkei NEEDS
tick archive~\cite{needs}.
The raw deliverable, however, is far from analysis-ready: a single year of
individual-stock ticks arrives as thousands of zipped CSV files, each trading
day split across numbered multi-part archives; four distinct data types share
the delivery with different schemas; the schemas themselves changed between
the 2016 and 2017 eras, with the 2016 index files using fixed-width rather
than CSV records; and column layouts are documented in Japanese, with
categorical fields encoded as numeric codes.
Researchers who license the data typically re-implement the same brittle
parsing, decoding, and filtering code before any economics can begin.

The cost of doing this in-house is easy to underestimate.
A working parser has to encode four file codes and two schema eras, a
fixed-width 2016 layout, 187 bilingual column names, and categorical fields
whose meanings live in Japanese-language vendor manuals---and the mistakes
surface late: the structured pre-release passes reported in
Appendix~\ref{app:validation} still found, in a mature version, a full-day
read that exhausted memory as an untrappable panic and a query that
truncated silently at its row cap.
The machine cost compounds it.
The pandas prototype this library replaced needed 286\,s for one of a
trading day's nine parts (Section~\ref{sec:performance}); at that rate a
single 2017-scale year of individual-stock ticks is on the order of a week
of wall-clock parsing, on an archive that has since grown from 101.9\,GB to
351.9\,GB a year.
What \texttt{tse\_tick} offers instead is convenience of a specific kind:
\texttt{pip install tse-tick}, and the four types, two eras, coded fields,
and Japanese layouts are already handled---in either language, behind two
calls.

\texttt{tse\_tick} closes this gap.
It is a Python library, built on Polars~\cite{polars} and Apache
Arrow~\cite{arrow}, that turns raw NEEDS archives into clean, explicitly
typed DataFrames with English or Japanese column names, and---for repeated
analysis---into a Hive-partitioned Parquet~\cite{parquet} store served by
embedded DuckDB~\cite{duckdb} SQL.
Its contributions are:
\begin{itemize}[nosep]
  \item a unified parser for all four NEEDS data types and both schema eras,
        with automatic type and era detection from file names
        (Section~\ref{sec:data});
  \item two access paths over one parse core---a one-shot filtered reader
        and a two-stage ingest-then-query pipeline---so that exploratory and
        production workflows use the same cleaning semantics
        (Section~\ref{sec:architecture});
  \item an ingestion engine designed for commodity hardware: per-date atomic
        units with resume safety, morsel-bounded streaming writes whose peak
        memory is independent of trading-day size, RAM-aware process
        parallelism, and part-pruning that exploits the archive's physical
        layout (Sections~\ref{sec:ingest}--\ref{sec:partscan});
  \item analysis utilities for order-book features and event-window
        extraction, plus a translation layer that maps familiar API names
        from yfinance, Polygon, and ccxt onto \texttt{tse\_tick} equivalents
        (Section~\ref{sec:query}, Appendix~\ref{app:translate});
  \item a reproducible benchmark suite quantifying the engine and storage
        choices (Section~\ref{sec:performance}).
\end{itemize}
The library supports data from 2016 through 2025, requires only Python
$\geq$3.9 with Polars and PyArrow (DuckDB is an optional extra), and is
installed with \texttt{pip install tse-tick}.
This paper describes version 0.15.1.

\section{The Nikkei NEEDS Data Format}
\label{sec:data}

\subsection{Data types and eras}

NEEDS tick deliveries comprise four data types, summarized in
Table~\ref{tab:datatypes}.
Individual-stock ticks (file code TICST120) are the largest by far: each
trading day is split into numbered ZIP parts ordered by ascending stock
code, plus a final ``appendix'' part, and a busy day can span dozens of
parts.
The other three types arrive as one monthly archive each.
Raw volumes grow steadily: the individual-stock archive for 2017 totals
101.9\,GB of compressed data and the 2025 archive 351.9\,GB, with the
largest single month (April 2025) at 42.0\,GB.
A complete 2016--2025 delivery of all four types is 35{,}080 archives and
1.725\,TiB, of which the individual-stock ticks alone are 34{,}720
archives and 1.72\,TiB---large enough that local storage, rather than
parsing, is often the binding constraint on a research machine.

\begin{table}[t]
\centering
\caption{The Four NEEDS Data Types Handled by \texttt{tse\_tick}.}
\label{tab:datatypes}
\small
\sbox{\tabbox}{%
\begin{tabular}{@{}llP{4.6cm}l@{}}
\toprule
\textbf{Data Type} & \textbf{File Code} & \textbf{Content} & \textbf{Output Fields (Raw)} \\
\midrule
\texttt{individual\_stock} & TICST120 & Trade-and-quote ticks per stock; ten quote levels per side & 95 \\
\texttt{stock\_summary} & TICSS110 & Daily per-stock summary records & 82 (83) \\
\texttt{indices} & TICIT110 & Intraday index updates & 10 (23; 15 in 2016) \\
\texttt{indices\_summary} & TICIS110 & Daily index summary records & 17 (83 from 2017) \\
\bottomrule
\end{tabular}}
\usebox{\tabbox}

\begin{tablenote}
\textit{Note:} Field counts follow the library's convention of reporting
cleaned \emph{output} columns, with raw parsed column counts in parentheses
where they differ. In 2016 the two index types were delivered as fixed-width
files under the legacy codes TICIT010 and TICIS010; from 2017 all types are
CSV. The era is detected automatically from the ZIP file name.
\end{tablenote}
\end{table}

The schema break between eras is not cosmetic.
The 2016 index-tick files use fixed-width 69-byte records with a 15-column
layout, whereas the 2017-onward files are CSV with 23 raw columns; the index
summary type similarly moved from a compact 2016 hybrid to an 83-column
summary layout.
\texttt{tse\_tick} detects the type and era from the file name and routes to
the appropriate parser, so user code is era-agnostic.
Table~\ref{tab:schema-summary} gives the shape of each type's layout at a
glance; complete column-level schemas for all four types, in both languages,
are reproduced in Appendix~\ref{sec:appendix-schema}.

\begin{table}[t]
\centering
\caption{Schema Summary by Data Type.}
\label{tab:schema-summary}
\small
\sbox{\tabbox}{%
\begin{tabular}{@{}lP{10.2cm}@{}}
\toprule
\textbf{Data Type (Output Fields)} & \textbf{Field Groups} \\
\midrule
\texttt{individual\_stock} (95) & identification 6 $\cdot$ timestamps 5 $\cdot$ trade 6 $\cdot$ best quote, both sides 6 $\cdot$ depth levels 2--10, both sides 54 $\cdot$ limit and special quotes 12 $\cdot$ OVER/UNDER overflow 6 \\
\addlinespace
\texttt{stock\_summary} (82) & identification 5 $\cdot$ share attributes 2 $\cdot$ execution-size histogram 8 $\cdot$ morning session 25 $\cdot$ afternoon session 25 $\cdot$ full-day aggregates 5 $\cdot$ quote-side execution splits 12 \\
\addlinespace
\texttt{indices} (10) & identification 6 $\cdot$ index tick fields 4 \\
\addlinespace
\texttt{indices\_summary} (17) & identification 5 $\cdot$ morning session 6 $\cdot$ afternoon session 6 \\
\bottomrule
\end{tabular}}
\usebox{\tabbox}

\begin{tablenote}
\textit{Note:} Group sizes are counts of cleaned output columns and sum to
each type's field count. The ten-level quote ladder dominates the
individual-stock layout, and the two summary types are session-blocked
rather than time-stamped. Column-by-column definitions in both languages
are in Appendix~\ref{sec:appendix-schema}.
\end{tablenote}
\end{table}

\subsection{Bilingual schemas and coded fields}

NEEDS documents its layouts in Japanese.
The library ships full English and Japanese column definitions for every
type---a 187-entry mapping connects the two, e.g.\ \emph{Stock Code}
$\leftrightarrow$ \mbox{銘柄コード} and \emph{Execution Price} $\leftrightarrow$
\mbox{約定価格}---and every reader accepts \texttt{language="en"} or
\texttt{"jp"}.
Categorical fields arrive as numeric codes and are decoded into readable
labels during cleaning; for example, the volume flag decodes
\texttt{"0"}~$\to$~\texttt{Final} and \texttt{"128"}~$\to$~\texttt{Estimated}.
Time fields are normalized to fixed-width \texttt{HHMMSS} strings across
types and eras.

\section{Related Work}
\label{sec:related}

Tooling for reconstructing research-grade microstructure data exists for
other venues.
LOBSTER~\cite{lobster} reconstructs limit-order books from NASDAQ ITCH
feeds and distributes ready-made event files widely used in the empirical
literature.
ArcticDB~\cite{arcticdb} provides a high-performance DataFrame store for
tick data, but is agnostic to any vendor's raw format---it solves storage,
not parsing.
Popular market-data client libraries such as yfinance, Polygon's official
client, and ccxt wrap vendor \emph{APIs} rather than local file deliveries,
and none understands the NEEDS layout.

Closed tooling for Japanese market data exists, but not for this
deliverable.
Nikkei's own retrieval front-end over the NEEDS databases, NEEDS
FinancialQUEST~\cite{fquest}, is a browser-based download tool for
fundamentals and daily price data rather than for the tick archive.
On the exchange side, JPX Market Innovation \& Research sells FLEX
Historical packet captures of the Tokyo Stock Exchange feed and, since June
2026, a processed ten-level order-book historical dataset~\cite{jpxflex};
both are commercial products built on the exchange's own messages rather
than on the NEEDS delivery.
Groups that license the NEEDS tick archive are left with in-house code.
\texttt{tse\_tick} is, to our knowledge, the first open-source library
specific to the NEEDS tick format; it complements rather than competes with
the tools above, and its translation layer (Appendix~\ref{app:translate})
deliberately bridges the naming conventions of those client libraries so
their users can locate the equivalent \texttt{tse\_tick} calls.

\section{Software Architecture}
\label{sec:architecture}

\subsection{Two access paths, one core}
\label{sec:twopaths}

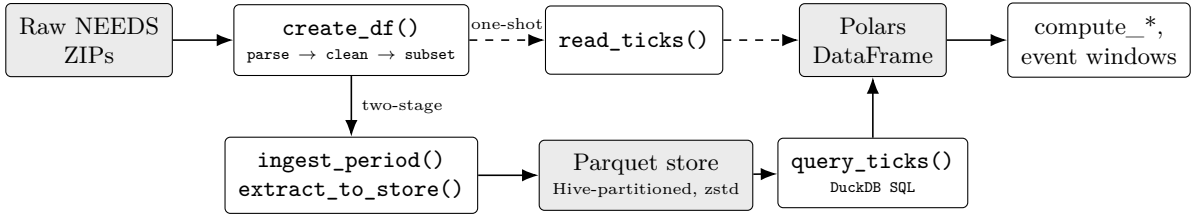
\begin{figure}[t]
\centering
\begin{tikzpicture}[
    font=\small,
    box/.style={draw, rounded corners=2pt, align=center, inner sep=5pt, minimum height=9mm},
    code/.style={box, font=\footnotesize\ttfamily},
    data/.style={box, fill=black!8, font=\footnotesize},
    arr/.style={-{Latex[length=2.2mm]}, semithick},
]
  \node[data] (zip) {Raw NEEDS\\ZIPs};
  \node[code, right=8mm of zip] (core) {create\_df()\\[-2pt]{\tiny parse $\to$ clean $\to$ subset}};
  \node[code, right=10mm of core] (oneshot) {read\_ticks()};
  \node[data, right=10mm of oneshot] (result) {Polars\\DataFrame};
  \node[box, right=8mm of result, font=\footnotesize] (analysis) {compute\_*,\\event windows};
  \node[code, below=8mm of core] (ingest) {ingest\_period()\\extract\_to\_store()};
  \node[data, right=8mm of ingest] (store) {Parquet store\\[-2pt]{\tiny Hive-partitioned, zstd}};
  \node[code, below=8mm of result] (query) {query\_ticks()\\[-2pt]{\tiny DuckDB SQL}};

  \draw[arr] (zip) -- (core);
  \draw[arr, dashed] (core) -- node[above, font=\tiny]{one-shot} (oneshot);
  \draw[arr, dashed] (oneshot) -- (result);
  \draw[arr] (core) -- node[right, font=\tiny]{two-stage} (ingest);
  \draw[arr] (ingest) -- (store);
  \draw[arr] (store) -- (query);
  \draw[arr] (query) -- (result);
  \draw[arr] (result) -- (analysis);
\end{tikzpicture}
\caption{The two access paths of \texttt{tse\_tick}. The dashed one-shot
path returns a filtered DataFrame directly from raw ZIPs; the two-stage
path ingests into a Hive-partitioned Parquet store served by DuckDB.
Both share the same parse-and-clean core.}
\label{fig:architecture}
\end{figure}

Every code path in \texttt{tse\_tick} funnels through a single
parse-and-clean core, exposed through two access paths
(Figure~\ref{fig:architecture}):

\begin{enumerate}[nosep]
  \item \textbf{One-shot}: \texttt{read\_ticks()} reads raw ZIPs straight
        into a filtered DataFrame with no intermediate store, selecting
        tickers, dates, and intraday time ranges through keyword filters.
        It caps results at 10~million rows by default (emitting a
        capturable \texttt{TruncationWarning}) and guards cumulative
        decompressed input at 5\,GB, raising a catchable
        \texttt{OneShotMemoryError} rather than exhausting RAM.
  \item \textbf{Two-stage}: the \texttt{ingest\_*} family converts archives
        into a Hive-partitioned Parquet store once; \texttt{query\_ticks}
        then serves repeated, sub-second filtered queries through DuckDB.
        \texttt{extract\_to\_store()} performs both stages in one call for
        a chosen ticker set and period.
\end{enumerate}

Table~\ref{tab:modules} lists the package modules.

\begin{table}[t]
\centering
\caption{Module Structure of \texttt{tse\_tick}.}
\label{tab:modules}
\small
\begin{tabular}{@{}lP{9.8cm}@{}}
\toprule
\textbf{Module} & \textbf{Responsibility} \\
\midrule
\texttt{enhanced.py} & One-shot core: \texttt{create\_df}, \texttt{read\_ticks}, \texttt{discover\_zips}, \texttt{parse\_period}, ZIP-bomb guards, warning and error types \\
\texttt{core.py} & \texttt{clean\_data} (casts and categorical decoding), 2016 fixed-width parser, tick timestamps \\
\texttt{schemas.py} & English and Japanese column definitions for the four types \\
\texttt{ingest.py} & Per-date grouped ingestion, process pool with RAM-aware worker cap, \texttt{extract\_to\_store} \\
\texttt{partscan.py} & Part-pruning: probe each part's first stock code, open only the ticker's run \\
\texttt{query.py} & DuckDB SQL over the Parquet store (optional \texttt{[query]} extra) \\
\texttt{event\_window.py} & Event-window extraction around corporate-event timestamps \\
\texttt{features.py} & Order-book features: spread, depth, flow imbalance, volatility \\
\texttt{io/parquet.py} & Hive-partitioned Parquet read and write \\
\texttt{constants.py} & \texttt{DataType} / \texttt{Language} enums and type-classification sets \\
\texttt{translate.py} & External-name translation (yfinance / Polygon / ccxt $\to$ \texttt{tse\_tick}) \\
\texttt{cli.py} & \texttt{tse-tick ingest} and \texttt{tse-tick export} command-line verbs \\
\bottomrule
\end{tabular}
\end{table}

\subsection{The parse-and-clean core}
\label{sec:parsecore}

NEEDS CSV content mixes numeric fields, zero-padded codes, blank-heavy
quote columns, and coded categoricals in a 95-column layout; naive type
inference fails unpredictably across files.
\texttt{create\_df} therefore reads \emph{every} column as a string, assigns
schema names, and only then applies explicit casts in \texttt{clean\_data}:
integer volume and flag columns to \texttt{Int64}, price and quote-price
columns to \texttt{Float64}, time strings normalized to \texttt{HHMMSS},
whitespace stripped vectorized, and categorical codes decoded in batched
expressions (batching the decode gave a measured $2.15\times$ on four
cores with byte-identical output).
Japanese output reuses the identical pipeline: columns are renamed
JP$\to$EN internally, cleaned in English, and renamed back, so both
languages share one tested code path.
Cleaned frames are subset to each type's documented output columns
(Table~\ref{tab:datatypes}).

\subsection{The ingestion engine}
\label{sec:ingest}

Ingestion converts raw archives into the store in per-date atomic units:
all ZIP parts of a trading day are processed together, so multi-part days
can never be half-written.
Resume safety is keyed per date through a coverage marker that records what
the partition actually contains, rather than inferring completion from file
existence---a store built for ticker~A must not cause a later request for
ticker~B to be skipped.
Partition writes are atomic (hidden temporary file, then an atomic rename),
and resume validates Parquet footers before trusting existing output.

For \texttt{individual\_stock} days with a ticker filter of at most 64
codes, the engine streams: each part is parsed and cleaned in
newline-aligned 64\,MB
\emph{morsels} (in the sense of morsel-driven processing~\cite{leis2014})
that are handed directly to a partitioned Parquet appender as row groups,
so neither a part's frame nor the day's frame is ever materialized.
Peak memory becomes independent of trading-day size: on the worst measured
real day, peak ingestion memory fell from 24.52\,GB to 2.40\,GB
($10\times$) with a byte-identical store.
Full-frame days and the summary and index types use a concatenate path
sized by a per-day estimate.

Date units are independent, so ingestion parallelizes across a
\texttt{spawn}-started process pool.
A RAM-aware cap bounds workers to the logical core count \emph{and} to what
fits in 70\% of available memory (measured as \texttt{MemAvailable} on
Linux and its Windows equivalent), with each worker's internal Polars
thread count bounded so the pool does not oversubscribe cores.
Measured on the reference 16-thread machine over one month of
ticker-filtered \texttt{individual\_stock} ingestion (19 trading days,
4.07~million kept rows, two interleaved repetitions), the pool reaches
$1.9\times$ at two workers and $3.3\times$ at four; a request for eight is
clamped to five by the memory ceiling---3.0\,GB per streaming worker
against 70\% of available RAM---and reaches $3.6\times$.
The store is byte-identical at every worker count.
Appendix~\ref{app:beta} reports the same shape of workload at production
scale on a 36-thread host.

\subsection{Part-pruning}
\label{sec:partscan}

NEEDS numbers a day's individual-stock parts in ascending stock-code order,
so any single ticker lives in a contiguous run of parts (plus the day's
final appendix part).
The part scanner probes only each part's first record, arithmetically
bounds the ticker's run, and opens just that run plus the last part---a
median $6.3\times$ faster single-ticker read (range $1.3$--$25.8\times$)
across 18 days sampled over three years, every one of them returning a
frame row-for-row identical to a full scan.
Whenever the ascending layout cannot be confirmed, the scanner falls back
to a full scan, so pruning is never less correct than scanning.
Stock codes are compared as fixed-width four-character tokens because the
TSE began assigning alphanumeric codes (e.g.\ \texttt{162A}) in 2024; on
one affected 27-part day, restoring pruning cut a single-ticker read from
472\,s to 13\,s.

\subsection{Store layout}
\label{sec:store}

The store is Hive-partitioned Parquet, zstd-compressed by default:
tick types are laid out as
\texttt{\{type\}/date=YYYYMMDD/ticker=CODE.parquet}, so DuckDB prunes both
date and ticker partitions before touching data.
The two summary types instead write one file per date with the code kept as
a column: per-ticker files would explode a 15\,MB month into tens of
thousands of near-empty files.
For the two index types the partition key is the raw numeric index code
(e.g.\ \texttt{101}), while query filters also accept display names.

Since version 0.15.0 every \texttt{individual\_stock} partition also
materializes an \texttt{Effective Time} key---an \texttt{Int32}
\texttt{HHMMSS} column resolving the trade-versus-quote timestamp
choice---so time-window predicates can use Parquet row-group statistics
instead of a computed expression that defeats them.
Measured on a real trading day's 2.56-million-row single-ticker
partition, a one-minute slice query
accelerated $7.64\times$ and a five-minute window $5.65\times$, for a
storage cost of $+0.52\%$; the key is excluded from documented query
output, so schemas are unchanged.

\section{Query and Analysis Layer}
\label{sec:query}

\subsection{Store queries}

\texttt{query\_ticks()} builds a pruned DuckDB scan over the store and
returns a Polars DataFrame, accepting a data type, ticker, date, intraday
time range, column subset, and row limit as keyword filters.
DuckDB is an optional dependency (\texttt{pip install "tse-tick[query]"});
the core install stays polars-plus-pyarrow.
Dates accept a day, a month, a year, or a range; four-character stock codes
select the parent listing together with its suffixed share classes, while
five-character codes select exactly one class.
Results are capped at 10~million rows with a \texttt{TruncationWarning}
(the cap probes \texttt{limit}+1, so truncation is detected, not guessed),
and zero-row results warn through a capturable \texttt{NoDataWarning} while
still returning a correctly typed empty frame.
For extracts beyond the cap, \texttt{export\_query} streams a store slice
to a single Parquet file with bounded memory, and
\texttt{extract\_to\_store} builds ticker-scoped stores without any row
cap.
\texttt{query\_sql} exposes raw SQL over the store as a documented escape
hatch, and \texttt{get\_available\_dates} / \texttt{get\_available\_tickers}
enumerate store contents.

\subsection{Optional analysis helpers}
\label{sec:helpers}

The two modules below are conveniences layered on query output rather than
parts of the parse-store-query core: each takes an ordinary DataFrame and
returns one, so a user with existing feature code can ignore them entirely
and lose nothing.

\paragraph{Order-book features.}
\texttt{features.py} computes standard microstructure quantities directly
on query output:
\begin{itemize}[nosep]
  \item \texttt{compute\_spread}: best bid--ask spread;
  \item \texttt{compute\_depth}: quote volumes for up to ten levels per side;
  \item \texttt{compute\_flow\_imbalance}: rolling signed-volume imbalance;
  \item \texttt{compute\_volatility}: rolling realized or Garman--Klass
        volatility over trade rows only, with \texttt{null}---not
        \texttt{NaN}---for undefined rows.
\end{itemize}
\texttt{compute\_all\_features} appends all of the above to the input
frame.

\paragraph{Event windows.}
\texttt{extract\_event\_window()} takes a ticker, an event date and time,
and window sizes (default $\pm 60$ minutes), and returns the ticks around
the event with a signed \texttt{seconds\_from\_event} column, falling back
to quote-update times for rows without an execution time; a batch variant
processes an events table.
A dedicated ingestion mode consumes an event list with precomputed reaction
anchors---for after-hours announcements the anchor is the next session's
09:00 open, since centering on the announcement itself would yield an empty
window---and writes a compact \texttt{year=YYYY/month=MM} store of window
slices; it is reachable as \texttt{ingest\_event\_windows\_period} and from
the command line as \texttt{tse-tick ingest --filter-csv}.

\subsection{The translation layer}

Researchers arriving from yfinance, Polygon, or ccxt can ask the library
for its equivalent of the API names they already know:
\texttt{translate("yfinance", "download")} returns
\texttt{"create\_df"}, and \texttt{mapping()} exposes the full table
programmatically.
The mapping ships as packaged JSON, can be extended through an environment
variable, and is reproduced in full in Appendix~\ref{app:translate}.

\section{Usage Examples}
\label{sec:usage}

Listing~\ref{lst:createdf} shows the core reader; Listing~\ref{lst:readticks}
the one-shot path; Listing~\ref{lst:twostage} the two-stage pipeline;
Listing~\ref{lst:features} the feature chain; Listing~\ref{lst:translate}
the enums and translation helpers; and Listing~\ref{lst:cli} the
command-line equivalents.

\begin{lstlisting}[caption={Parsing a raw ZIP into a typed DataFrame.},label={lst:createdf}]
import tse_tick as tt

df    = tt.create_df("HTICST120.20230704.1.zip", language="en")  # 95 columns
df_jp = tt.create_df("HTICST120.20230704.1.zip", language="jp")  # Japanese names
head  = tt.create_df("HTICST120.20230704.1.zip", rows=10)        # capped preview
\end{lstlisting}

\begin{lstlisting}[caption={One-shot reading: raw ZIPs to a filtered DataFrame, no store.},label={lst:readticks}]
df = tt.read_ticks(
    r"G:\data\needs",          # folder of raw ZIPs (or a single ZIP)
    ticker_filter="7203",      # Toyota; also accepts iterables of codes
    date="20230704",
    start_time="09:00:00",
    end_time="11:30:00",
)
\end{lstlisting}

\begin{lstlisting}[caption={Two-stage pipeline: build a ticker-scoped store once, query repeatedly.},label={lst:twostage}]
from tse_tick import extract_to_store, query_ticks

extract_to_store(r"G:\data\needs", "store", period="202301-202312",
                 ticker="7203", max_workers="auto")

df = query_ticks("store", data_type="individual_stock", ticker=7203,
                 date="20230704", start_time="09:00:00", end_time="11:30:00")
\end{lstlisting}

\begin{lstlisting}[caption={Order-book features on query output.},label={lst:features}]
from tse_tick import (compute_spread, compute_depth, compute_flow_imbalance,
                      compute_volatility, compute_all_features)

df       = query_ticks("store", ticker=7203, date="20230704")
spread   = compute_spread(df)
depth    = compute_depth(df, levels=5, side="both")
ofi      = compute_flow_imbalance(df, window="5min")
vol      = compute_volatility(df, window="5min")
features = compute_all_features(df)   # input columns + 23 feature columns
\end{lstlisting}

\begin{lstlisting}[caption={Enums and the name-translation layer.},label={lst:translate}]
from tse_tick import DataType, Language, translate, mapping

DataType.INDIVIDUAL_STOCK                     # 'individual_stock' (str enum)
Language.JP                                   # 'jp'
translate("yfinance", "download")             # -> 'create_df'
translate("polygon", "from_")                 # -> 'start_time'
mapping("ccxt")["functions"]["fetch_trades"]  # -> ['query_ticks', 'read_ticks']
\end{lstlisting}

The same workflows are scriptable without Python through the CLI:

\begin{lstlisting}[language={},caption={Command-line interface.},label={lst:cli}]
tse-tick ingest --data-type individual_stock --period 202301-202312 \
         --input-root G:\data\needs --output-root store \
         --tickers 7203 --parallel auto

tse-tick export --data-type individual_stock --input-root G:\data\needs \
         --period 20230704 --tickers 7203 --output toyota_20230704.csv
\end{lstlisting}

\noindent
\texttt{ingest} accepts periods from a single day to a multi-year range,
resumes interrupted runs by default, and sizes its worker pool
automatically; \texttt{export} writes CSV or Parquet chosen by the output
extension, and \texttt{--store} switches it to the two-stage path.

\input{performance_section.tex}

\section{Testing}
\label{sec:testing}

The test suite collects 637 tests and fails nowhere on either supported
platform.
Because the underlying data is proprietary, the suite is synthetic-first: a
fixture generator emits obviously fake NEEDS-format archives that are
pushed through the \emph{real} ingest pipeline, so parsing, cleaning,
storage, query, feature, and CLI behavior are all exercised without any
licensed data.
The generator emits individual-stock, stock-summary, and index
archives---the last in both the 2017-onward CSV and the fixed-width 2016
layouts---so era-specific parsing is exercised on synthetic input as well.
Tests that require real archives are gated on their presence
(Table~\ref{tab:tests}).
Three conventions keep the suite honest as the library changes: every fixed
defect keeps its own regression file---\texttt{test\_alpha\_fixes.py} and
one per audit round---so a bug that has been fixed once cannot quietly
return; every performance change proves output identity, a byte-identical
store or a row-identical frame, alongside its speedup, which is the
evidence on which the streaming write path and part-pruning were accepted;
and drift tests assert that data-type classification stays single-sourced,
so the four types cannot come to disagree between modules.
Continuous integration runs Python 3.9, 3.11, and 3.13; the usage listings
in Section~\ref{sec:usage} are themselves locked by tests
(\texttt{tests/test\_paper\_examples.py}), so the printed API cannot drift
from the shipped one.
Beyond the automated suite, two structured pre-release passes were run on
machines other than the development host---an alpha pass across all four
data types, both schema eras, and both access paths, and an independent
beta pass over a three-year ingestion---and are reported in
Appendix~\ref{app:validation}.

\begin{table}[t]
\centering
\caption{Test-Suite Profiles (637 Collected, Version 0.15.1).}
\label{tab:tests}
\small
\begin{tabular}{@{}lll@{}}
\toprule
\textbf{Environment} & \textbf{Result} & \textbf{Skips} \\
\midrule
Without licensed data (any OS) & 587 pass / 50 skip & 48 data-gated, 2 platform-gated \\
Windows, with licensed data & 635 pass / 2 skip & case-sensitive-filesystem tests \\
Linux, with licensed data & 636 pass / 1 skip & Windows-console encoding test \\
\bottomrule
\end{tabular}
\end{table}

\section{Security and Robustness}
\label{sec:security}

The library treats untrusted archives and user-supplied query parameters
defensively; Table~\ref{tab:guards} summarizes the guards, whose values are
deliberate design decisions rather than defaults.

\paragraph{Archives are validated before they are decompressed.}
Every entry is checked against a decompressed-size ceiling, a compression
ratio, and an entry count before any bytes are expanded, and a violation
raises \texttt{SuspiciousZipError} instead of filling memory or disk.
A NEEDS archive holds a single CSV, so a five-entry ceiling is generous for
legitimate input while still refusing the nesting a ZIP bomb needs.

\paragraph{Row and memory limits are explicit, and refusals happen up front.}
Reads are capped at 10~million rows.
In the query path the cap is enforced by asking for one row beyond the
limit, so a truncated result is detected rather than guessed at from an
exactly-full one; the one-shot reader additionally tracks whether any
archive was left unread, and warns on that rather than on the row count
alone.
That path also sums the decompressed size it is about to materialize and
refuses the read with a catchable
\texttt{OneShotMemoryError} before allocating; the query path raises
\texttt{QueryMemoryError} on the same principle.
Callers who need more rows than the cap allows use the paths built to
stream---\texttt{export\_query} for a single file,
\texttt{extract\_to\_store} for a ticker-scoped store---neither of which is
capped.

\paragraph{User input reaching SQL or the filesystem is validated, not trusted.}
Identifiers interpolated into generated SQL pass a character blocklist that
rejects quoting and control characters while still allowing the spaces that
legitimate NEEDS column names contain---which is why the check is a
blocklist rather than an alphanumeric allowlist.
Dates, times, and ticker codes are matched against strict patterns before
they reach a path or a predicate.

\paragraph{Failures are typed values, not printed text.}
Empty results warn through \texttt{NoDataWarning} and still return a
correctly typed zero-row frame; truncation warns through
\texttt{TruncationWarning}; a failed ingest worker surfaces as
\texttt{IngestWorkerError} rather than as a silent gap in the store.
All of them are capturable, so a batch job can decide for itself what a
warning means.
Library code logs and never prints---only the CLI prints---and importing
the package reconfigures the process's standard streams to UTF-8, so a
Japanese Windows console cannot turn non-ASCII output into an encoding
error.

\begin{table}[t]
\centering
\caption{Resource and Input Guards.}
\label{tab:guards}
\small
\begin{tabular}{@{}lP{6.6cm}@{}}
\toprule
\textbf{Guard} & \textbf{Default} \\
\midrule
ZIP decompressed size & 5\,GB per ZIP entry \\
ZIP compression ratio & 100:1 maximum \\
ZIP entry count & 5 entries maximum \\
One-shot decompressed input & 5\,GB (\texttt{max\_oneshot\_bytes}, catchable) \\
Query / one-shot row cap & 10,000,000 rows, warning on truncation \\
Ingest worker pool & RAM-aware cap: cores $\wedge$ 70\% of available RAM \\
SQL identifiers & Character blocklist; strict date, time, ticker patterns \\
\bottomrule
\end{tabular}
\end{table}

\section{Limitations}
\label{sec:limitations}

\texttt{tse\_tick} ships no data---NEEDS archives require an institutional
license---and parses that format alone, though its Parquet stores are
readable by any Arrow-compatible tool. Four behaviors are deliberate:
same-timestamp tie order is unfixed, so exports of one slice on different
platforms are multiset-equal but not byte-identical; \texttt{read\_ticks}
and \texttt{query\_ticks} cap results at 10~million rows, with
\texttt{extract\_to\_store} and \texttt{export\_query} uncapped; intraday
time filters are rejected for the daily \texttt{*\_summary} types with a
clear error; and event-window ingestion consumes precomputed reaction
anchors rather than deriving them from an exchange calendar.

\paragraph{Maintenance and schema evolution.}
The library describes the 2016--2025 delivery, and data formats outlive
papers.
Releases follow semantic versioning, and the store schema has changed twice
so far (in versions 0.4.0 and 0.9.0), each recorded in the changelog
together with the re-ingest it implies.
Era detection is filename-driven and additive, so a post-2025 layout enters
as another branch of the same detector plus its synthetic fixtures, without
touching the code paths of existing eras---and because the test suite is
synthetic-first, that work needs no licensed data: a contributor holding a
new delivery can encode it from the vendor manual alone.
Development continues in the open at
\url{https://github.com/tse-tick/tse_tick}, where issues and pull requests
are the supported channel, and the authors intend to track NEEDS format
changes for as long as the data underpins their own research.

\section{Conclusion}
\label{sec:conclusion}

\texttt{tse\_tick} turns the raw Nikkei NEEDS tick delivery---four data
types, two schema eras, thousands of multi-part Japanese-documented
archives---into typed DataFrames and a portable Parquet store on commodity
hardware.
One parse core serves both an exploratory one-shot reader and a production
two-stage pipeline whose ingestion is atomic, resumable, memory-bounded,
and RAM-aware parallel.
It is MIT-licensed, tested across platforms and Python versions, and
available at \url{https://github.com/tse-tick/tse_tick}.

\section*{Acknowledgments}
We thank Yutaka Ozeki (Nakatsuma Seminar, Keio University) for independently beta-testing version 0.15.1 against the 2023–2025 archives.

\section*{AI-Assisted Writing Disclosure}

Portions of this manuscript and of the described software were drafted with
the assistance of AI coding and writing tools (GitHub Copilot, OpenAI
Codex, xAI Grok, and Anthropic Claude).
All AI-assisted content was reviewed, verified, and edited by the authors,
who take responsibility for the final text and software.

% The bibliography alone is set ragged-right. Several entries end in long
% service URLs that cannot break tightly enough to justify, which leaves the
% preceding line loose past \hbadness; ragged-right removes the stretch
% entirely rather than trading a citation's precision for a tidy margin.
{\raggedright
\bibliographystyle{plain}
\bibliography{main}
\par}

\appendix
\clearpage

% Appendix tables are numbered A.1, B.1, C.1 ... so a cross-reference says at a
% glance whether it points into the main text or into the reference material.
% \@addtoreset restarts the counter in each appendix; both are package-free.
\makeatletter
\@addtoreset{table}{section}
\makeatother
\renewcommand{\thetable}{\thesection.\arabic{table}}
\setcounter{table}{0}

\noindent\textbf{The appendices are reference material.}
Appendix~\ref{app:validation} records the pre-release validation passes,
Appendix~\ref{app:translate} reproduces the complete API translation table,
and Appendix~\ref{sec:appendix-schema} the complete column-level schemas of
all four data types.
None of them is required to follow the main text; the schema tables are
intended for readers working directly against a NEEDS delivery.

\section{Pre-Release Validation}
\label{app:validation}

\subsection{Alpha Test (Version 0.11.4)}
\label{app:alpha}

One of the authors (M.~Hayashi) alpha-tested version 0.11.4, installing it
with its \texttt{[query]} extra from PyPI into a fresh Windows~11
environment on a Python release outside the project's CI matrix (3.14,
against CI's 3.9, 3.11 and 3.13), with no build step required.
All four data types parsed on both schema eras and both access paths;
one-shot and two-stage row counts agreed on real data; record counts and
dates matched the raw source; and full-width text was handled without
corruption.
With 46\,GB of free disk against the 1.725\,TiB corpus, slices were fetched,
verified, tested and deleted in turn rather than mirrored.
Three defects surfaced, all fixed in version 0.11.5 three days later
(Table~\ref{tab:alpha}), each with a regression test
(\texttt{tests/test\_alpha\_fixes.py}).

\begin{table}[ht]
\centering
\caption{Alpha-Test Findings and Their Resolution.}
\label{tab:alpha}
\small
\begin{tabular}{@{}P{6.2cm}P{7.4cm}@{}}
\toprule
\textbf{Finding (0.11.4)} & \textbf{Resolution (0.11.5)} \\
\midrule
A one-shot read of a full multi-part \texttt{individual\_stock} day
exhausted memory and surfaced as an uncatchable Rust panic, which
\texttt{except Exception} cannot trap &
A cumulative decompressed-size ceiling refuses the read up front, and any
panic during the load is converted, so both raise the catchable
\texttt{OneShotMemoryError} of Section~\ref{sec:twopaths} \\
\addlinespace
\texttt{query\_ticks} truncated silently at its row cap, unlike
\texttt{read\_ticks} &
The cap now emits \texttt{TruncationWarning}, distinguishing a truncated
result from an exactly-full one by probing one row beyond the limit
(Section~\ref{sec:query}) \\
\addlinespace
An explicit \texttt{year=} argument was ignored under automatic detection &
Detection now applies only to the arguments left unset, so an explicit
value is honored \\
\bottomrule
\end{tabular}
\end{table}

Two observations were recorded without code changes: stock codes are kept
exactly as delivered rather than normalized from full-width forms---none
has appeared in the archives sampled, so the case stays
latent---and a Japanese Windows console could raise an encoding error on
non-ASCII output, which the package now avoids by reconfiguring its
streams to UTF-8 at import.
The pass predates this release by four minor versions, but it is where the
one-shot memory guard and the truncation warning originate.

\subsection{Independent Beta Test (Version 0.15.1)}
\label{app:beta}

Version 0.15.1 was beta-tested independently by Yutaka Ozeki against the raw NEEDS archives for 2023--2025 on
the workstation of Table~\ref{tab:betaenv}.
The test was clean-slate: the prior dataset was wiped, the package updated
from PyPI, and a three-year single-ticker ingestion (Toyota, code 7203) run
with the native \texttt{max\_workers="auto"} selection, followed by a
whole-corpus time-window query.

\begin{table}[ht]
\centering
\caption{Beta-Test Environment.}
\label{tab:betaenv}
\small
\begin{tabular}{@{}ll@{}}
\toprule
\textbf{Component} & \textbf{Specification} \\
\midrule
CPU & Intel Core i9-9980XE, 18 cores / 36 threads \\
Memory & 128\,GiB total ($\approx$113\,GB available at launch) \\
Operating system & Windows 10 Pro (22H2) \\
Storage & Local NTFS volumes \\
Package & \texttt{tse-tick} 0.15.1 from PyPI \\
\bottomrule
\end{tabular}
\end{table}

\paragraph{Ingestion.}
All three years completed cleanly: 246 trading days for 2023, 241 for
2024, and 243 for 2025---730 in total---ingesting approximately
414~million rows in 2{,}104\,s ($\approx$35~minutes), i.e.\ roughly
2.9\,s per trading day or 197{,}000 rows per second end-to-end.

\paragraph{The RAM guard fired correctly.}
Automatic selection detected all 36 logical processors, applied the
streaming-path estimate of 3.0\,GB per worker (Section~\ref{sec:ingest}),
and found that 36 workers ($108$\,GB) would exceed the 70\% ceiling on
available memory ($\approx$79\,GB of 113\,GB); it clamped the pool to 26
($78$\,GB) before starting, and ingestion then ran without memory
exhaustion on precisely the kind of high-core, high-RAM host where an
unguarded pool would oversubscribe.

\paragraph{Whole-corpus window query.}
A five-minute market-open window (09:00:00--09:05:00) was then extracted
across the full three-year corpus by scanning all 730 daily Parquet
partitions with Polars directly, returning 22.3~million rows in 7.39\,s.
The store carried the materialized \texttt{Effective Time} key of
Section~\ref{sec:store}, so the bounded window could prune row groups, and
using Polars rather than the DuckDB layer exercised the store's
portability to other Arrow-compatible engines.

\paragraph{Scope.}
The workload is single-ticker, on hardware unlike the reference machine of
Section~\ref{sec:performance}: its figures, taken from the tester's logs,
attest guard-respecting operation and end-to-end usability rather than
being comparable to the controlled benchmarks.

\section{The API Translation Mapping}
\label{app:translate}

Table~\ref{tab:translate} reproduces the complete translation table the
package ships as \texttt{translations.json} (23 entries across three
source libraries).
Lookups resolve argument names before function names and exact-case matches
before case-insensitive ones; when one external call corresponds to several
\texttt{tse\_tick} calls, \texttt{translate()} returns the first listed and
\texttt{mapping()} exposes all of them.
Users can extend or override the table by pointing the
\texttt{TSE\_TICK\_TRANSLATIONS} environment variable at a JSON file with
the same shape.

% A longtable, not a float: the mapping is a reference listing at the end of a
% short appendix, so letting it break across the page boundary keeps the
% appendix flowing instead of stranding it on a float page of its own.
{\small
\setlength{\LTcapwidth}{\linewidth}
\begin{longtable}{@{}lllP{4.9cm}@{}}
\caption{Complete API Translation Mapping (Version 0.15.1).} \label{tab:translate} \\
\toprule
\textbf{Kind} & \textbf{External Name} & \textbf{\texttt{tse\_tick} Name} & \textbf{Note} \\
\midrule
\endfirsthead
\multicolumn{4}{@{}l}{\small\itshape Table~\ref{tab:translate} continued} \\
\toprule
\textbf{Kind} & \textbf{External Name} & \textbf{\texttt{tse\_tick} Name} & \textbf{Note} \\
\midrule
\endhead
\midrule
\multicolumn{4}{r@{}}{\small\itshape Continued on next page} \\
\endfoot
\bottomrule
\endlastfoot
\multicolumn{4}{@{}l}{\textbf{yfinance}} \\
function & \texttt{download} & \texttt{create\_df} & also \texttt{read\_ticks}, \texttt{ingest\_period} \\
function & \texttt{Ticker.history} & \texttt{query\_ticks} & \\
function & \texttt{history} & \texttt{query\_ticks} & \\
function & \texttt{tickers} & \texttt{get\_available\_tickers} & \\
function & \texttt{Tickers} & \texttt{get\_available\_tickers} & \\
argument & \texttt{tickers} & \texttt{ticker\_filter} & arguments resolve before functions \\
argument & \texttt{start} & \texttt{start\_time} & \\
argument & \texttt{end} & \texttt{end\_time} & \\
\addlinespace
\multicolumn{4}{@{}l}{\textbf{Polygon}} \\
function & \texttt{get\_aggs} & \texttt{query\_ticks} & \\
function & \texttt{list\_aggs} & \texttt{query\_ticks} & \\
function & \texttt{list\_trades} & \texttt{query\_ticks} & \\
function & \texttt{list\_tickers} & \texttt{get\_available\_tickers} & \\
argument & \texttt{ticker} & \texttt{ticker} & \\
argument & \texttt{from\_} & \texttt{start\_time} & \\
argument & \texttt{to} & \texttt{end\_time} & \\
argument & \texttt{limit} & \texttt{limit} & \\
\addlinespace
\multicolumn{4}{@{}l}{\textbf{ccxt}} \\
function & \texttt{fetch\_ohlcv} & \texttt{query\_ticks} & \\
function & \texttt{fetch\_trades} & \texttt{query\_ticks} & also \texttt{read\_ticks} \\
function & \texttt{symbols} & \texttt{get\_available\_tickers} & \\
function & \texttt{load\_markets} & \texttt{get\_available\_tickers} & \\
argument & \texttt{symbol} & \texttt{ticker} & \\
argument & \texttt{since} & \texttt{start\_time} & \\
argument & \texttt{limit} & \texttt{limit} & \\
\end{longtable}
}

\input{full_schema.tex}

\end{document}

%% file: performance_section.tex
\section{Performance Evaluation}
\label{sec:performance}

To validate the design choices described above, we benchmarked the
processing pipeline on representative files for all four NEEDS data
types from January~2017.
The headline comparison uses HTICST120 (individual stock ticks,
4.8\,million rows, 95~columns).
Two pandas~\cite{mckinney2010} baselines are reported: the
\emph{original prototype}, which
uses the Python CSV engine (\texttt{engine='python'}) because the NEEDS
CSV has ragged lines that the C engine rejects without additional
configuration; and a \emph{fair} baseline that uses the pandas C engine
with forced column count and all-string types (matching the Polars
parsing strategy), so that only the dataframe engine differs.

Table~\ref{tab:engine-benchmark} shows the results.
With all 16~logical cores, Polars completes parse-and-clean in
a median of 4.8\,s---three speedup comparisons tell different
parts of the story:
\begin{enumerate}[nosep]
  \item \emph{vs.\ the original prototype} (Python CSV engine,
        286\,s): $59.8\times$.
        This is what the rewrite achieved over the tool researchers
        previously used, but it conflates the CSV-parser change
        (Python~$\to$~C/Rust) with the dataframe-engine change.
  \item \emph{vs.\ the fair pandas baseline, 16 threads}
        (C engine, 164\,s): $34.3\times$.
        Both libraries now parse identically configured CSV; the
        gap is the dataframe engine plus Polars' multithreaded I/O.
  \item \emph{vs.\ the fair pandas baseline, single thread}
        (C engine, 164\,s; Polars 1-thread 22.8\,s): $7.2\times$.
        This is the purest library-vs-library comparison: single
        thread, same CSV configuration, same downstream work.
\end{enumerate}
\noindent The gap between the 16-thread ($34.3\times$) and 1-thread
($7.2\times$) factors reflects the parallelism that Polars exploits
internally; pandas' C engine is single-threaded, so its time is
the same in both comparisons.
Peak process memory is 13.1\,GiB for Polars, versus
28.1\,GiB for the prototype and 9.7\,GiB for the fair baseline.
All three figures are for a single monolithic parse of the whole part---what
the one-shot reader does without a ticker filter, and the only shape in which
the three engines are comparable---not for the streaming ingest path of
Section~\ref{sec:ingest}, whose peak is bounded per morsel instead.

The same benchmark was applied to the three remaining data types.
Fair-baseline speedups for Polars (16~threads):
\begin{itemize}[nosep]
  \item Stock Summary (HTICSS110, 82K rows, 82 cols (83 raw)): 2.1$\times$.
        The file is small enough that fixed per-file costs and CSV
        parsing damp the achievable gap.
  \item Index Ticks (HTICIT110, 1.2M rows, 10 cols (23 raw)): 7.1$\times$.
\end{itemize}
\noindent HTICIS110 (index summary) contains only 209~rows; at that
scale, fixed interpreter and library startup costs dominate, and Polars
is slower than pandas.
Its results appear in Table~\ref{tab:engine-benchmark} with a footnote
but are excluded from aggregate speedup claims.
The suite behind every number and figure in this section ships with the
library as \texttt{benchmarks/}, driven by \texttt{run\_all.py}.
It reads a local NEEDS mirror, so it is reproducible by any licensee.
A correctness gate in the same suite checks that the pandas and Polars
pipelines emit the same cleaned output for all four data types---shape,
integer and string equality, floats within $10^{-6}$---so the comparisons
are like-for-like rather than merely similarly named.

\input{engine_benchmark.tex}

Table~\ref{tab:format-comparison} evaluates storage formats.
Parquet with zstd compression---the store default---is
34.4$\times$ smaller than uncompressed CSV (68.9\,MB versus
2.37\,GB), yet reads the full file back in 0.29\,s versus 24.7\,s.
The columnar advantage is sharpest in selective reads: loading 3 of
95~columns from Parquet takes 0.0124\,s versus
12.42\,s for CSV
(1002$\times$ faster).
Snappy compression trades roughly twice the file size (139.9\,MB)
for a marginally faster write; Feather (Arrow~IPC) writes fastest
but is 4.8$\times$ larger than Parquet--zstd, reads more slowly,
and lacks the predicate pushdown and Hive partition pruning that
Parquet provides.
CSV.gz is smaller still (49.9\,MB), but pays for it on every read:
30.4\,s for the full file and 13.3\,s for the three-column read,
$105\times$ and $1{,}070\times$ slower than Parquet--zstd.

\input{format_comparison.tex}

\begin{figure}[t]
\centering
\includegraphics[width=\textwidth]{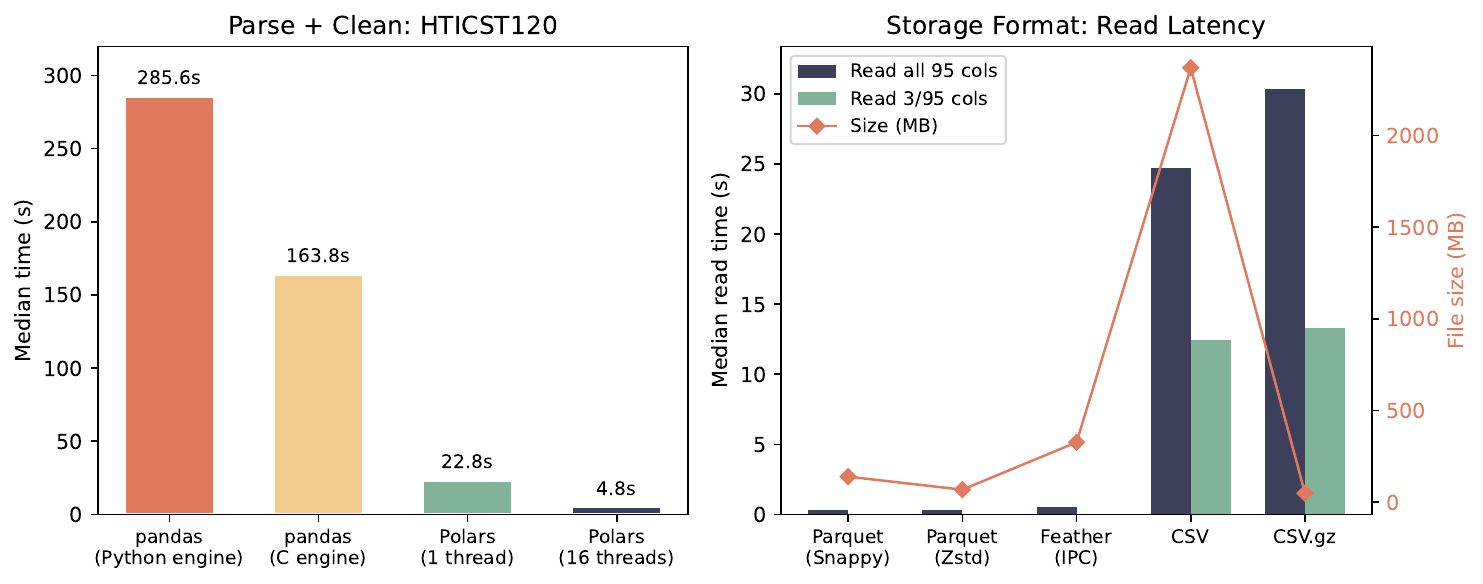}
\caption{Left: parse-and-clean time by processing engine (HTICST120).
Right: storage-format read latency and file size.}
\label{fig:benchmark}
\end{figure}

For the query workload central to \texttt{tse\_tick}'s design
(Table~\ref{tab:query-benchmark}),
a single-ticker one-hour slice from a Hive-partitioned Parquet
store (1,500,000~rows, a ticker subset of multi-day data)
completes in 0.0169\,s via DuckDB,
versus 6.92\,s for a pandas scan of the
equivalent monolithic CSV (410$\times$ faster).
DuckDB prunes irrelevant date and ticker partitions before
reading, touching only the rows that match the filter.
Figure~\ref{fig:benchmark} summarizes these results.

%% file: engine_benchmark.tex
\begin{table}[t]
\centering
\caption{Processing Engine Benchmark Across Four NEEDS Data Types (January 2017).}
\label{tab:engine-benchmark}
\small
% 7 columns at \small need tighter separation to hold \textwidth; \extracolsep
% equalises the visible gaps, so this is invisible next to Tables 4 and 5.
\setlength{\tabcolsep}{2pt}
\sbox{\tabbox}{%
\begin{tabular*}{\textwidth}{@{\extracolsep{\fill}}llrrrrr@{}}
\toprule
\textbf{Data Type} & \textbf{Backend} & \textbf{Rows} & \textbf{Median (s)} & \textbf{RSS (MiB)} & \textbf{vs.\ Proto.} & \textbf{vs.\ Fair} \\
\midrule
Individual Stock Ticks & pandas, Python eng. & 4,777,240 & 285.552 & 28783 & 1.0$\times$ & 0.6$\times$ \\
 & pandas, C eng. &  & 163.829 & 9967 & 1.7$\times$ & 1.0$\times$ \\
 & Polars, 16T &  & 4.774 & 13460 & 59.8$\times$ & 34.3$\times$ \\
 & Polars, 1T &  & 22.789 & 13059 & 12.5$\times$ & 7.2$\times$ \\
\addlinespace
Stock Summary & pandas, Python eng. & 81,688 & 3.785 & 625 & 1.0$\times$ & 0.5$\times$ \\
 & pandas, C eng. &  & 1.842 & 464 & 2.1$\times$ & 1.0$\times$ \\
 & Polars, 16T &  & 0.861 & 379 & 4.4$\times$ & 2.1$\times$ \\
 & Polars, 1T &  & 1.087 & 315 & 3.5$\times$ & 1.7$\times$ \\
\addlinespace
Index Ticks & pandas, Python eng. & 1,213,984 & 10.910 & 1462 & 1.0$\times$ & 0.6$\times$ \\
 & pandas, C eng. &  & 6.208 & 1067 & 1.8$\times$ & 1.0$\times$ \\
 & Polars, 16T &  & 0.880 & 1215 & 12.4$\times$ & 7.1$\times$ \\
 & Polars, 1T &  & 1.980 & 1190 & 5.5$\times$ & 3.1$\times$ \\
\addlinespace
Index Summary$^{\dagger}$ & pandas, Python eng. & 209 & 0.223 & 116 & 1.0$\times$ & 1.0$\times$ \\
 & pandas, C eng. &  & 0.217 & 117 & 1.0$\times$ & 1.0$\times$ \\
 & Polars, 16T &  & 0.636 & 132 & 0.4$\times$ & 0.3$\times$ \\
 & Polars, 1T &  & 0.617 & 125 & 0.4$\times$ & 0.4$\times$ \\
\bottomrule
\end{tabular*}}
\usebox{\tabbox}

\begin{tablenote}
\textit{Note:} Each condition was run 7~times in an isolated subprocess; the first run (warm-up) was discarded. Memory measured as peak process working set via \texttt{psutil}. ``Proto.''\ = original pandas prototype (\texttt{engine='python'}); ``Fair''\ = pandas with C engine, forced columns, all-string dtype. System: Intel Core i5-14400F (10-core/16-thread), 32\,GB RAM, Python~3.11, Polars~1.42, pandas~2.3, PyArrow~24.0, DuckDB~1.5.2; measured on \texttt{tse\_tick}~0.15.1 (2026-07-18 run). On the headline comparison the six measured runs span 4.723--4.785\,s (Polars, 16T) against 154.98--173.19\,s (fair baseline), so the median $34.3\times$ is bounded by $32.4\times$ and $36.7\times$ across the observed extremes.
\par\smallskip
$^{\dagger}$HTICIS110 has only 209~rows; at this scale fixed startup costs dominate and speedup ratios are not meaningful for throughput comparison.
\end{tablenote}
\end{table}

%% file: format_comparison.tex
\begin{table}[t]
\centering
\caption{Storage Format Comparison for HTICST120 (4.8\,M rows, 95 columns).}
\label{tab:format-comparison}
\small
\sbox{\tabbox}{%
\begin{tabular*}{\textwidth}{@{\extracolsep{\fill}}lrrrr@{}}
\toprule
\textbf{Format} & \textbf{Size (MB)} & \textbf{Write (s)} & \textbf{Read All (s)} & \textbf{Read 3/95 (s)} \\
\midrule
CSV & 2369.27 & 101.1405 & 24.6933 & 12.4232 \\
CSV.gz & 49.88 & 170.0348 & 30.3507 & 13.2691 \\
Parquet (Snappy) & 139.85 & 0.7107 & 0.3162 & 0.0146 \\
Parquet (Zstd) & 68.86 & 0.9116 & 0.2887 & 0.0124 \\
Feather (IPC) & 327.19 & 0.5827 & 0.5104 & 0.0608 \\
Pickle & 3202.08 & 14.8280 & 3.5429 & --- \\
\bottomrule
\end{tabular*}}
\usebox{\tabbox}

\begin{tablenote}
\textit{Note:} Medians of 5~runs (1~warm-up discarded). ``Read 3/95'' reads only Stock~Code, Execution~Price, and Volume. Pickle lacks selective reads. HDF5 omitted (pytables not installed).
\end{tablenote}
\end{table}

\begin{table}[t]
\centering
\caption{Query Latency: Single-Ticker Hour Slice from Multi-Day Store.}
\label{tab:query-benchmark}
\small
\begin{tabular*}{\textwidth}{@{\extracolsep{\fill}}lrrr@{}}
\toprule
\textbf{Method} & \textbf{Median (s)} & \textbf{Rows Returned} & \textbf{Speedup} \\
\midrule
DuckDB + Hive Parquet & 0.0169 & 52 & 409.6$\times$ \\
DuckDB + CSV scan & 0.8123 & 52 & 8.5$\times$ \\
pandas CSV scan & 6.9223 & 52 & 1.0$\times$ \\
\bottomrule
\end{tabular*}
\end{table}

%% file: full_schema.tex
\section{Complete Schema Documentation}
\label{sec:appendix-schema}

Tables~\ref{tab:ticst120-full}--\ref{tab:ticis110-full} give the four types'
complete schemas: column numbers are positions in the output DataFrame, the
Type column is each field's logical type (physical dtypes follow
Section~\ref{sec:parsecore}), and both name languages appear, selected with
\texttt{language}.

% ---- TICST120: full 95-column table ----
\begin{landscape}
    \setlength{\LTcapwidth}{\linewidth}
    \footnotesize
    \begin{longtable}{@{}p{0.7cm} P{4.0cm} P{4.2cm} p{1.0cm} P{9.0cm}@{}}
        \caption{TICST120 Individual Stock Tick Data Schema (95 fields).} \label{tab:ticst120-full}                                                            \\
        \toprule
        \textbf{Col} & \textbf{English Name} & \textbf{Japanese Name} & \textbf{Type} & \textbf{Description}                                                   \\
        \midrule
        \endfirsthead
        \multicolumn{5}{@{}l}{\small\itshape Table~\ref{tab:ticst120-full} continued}                                                                             \\
        \toprule
        \textbf{Col} & \textbf{English Name} & \textbf{Japanese Name} & \textbf{Type} & \textbf{Description}                                                   \\
        \midrule
        \endhead
        \midrule
        \multicolumn{5}{r@{}}{\small\itshape Continued on next page}                                                                                              \\
        \endfoot
        \bottomrule
        \endlastfoot
        1            & Record Type           & レコード種別                 & string        & Record format identifier (e.g., 1200 = Stocks with multiple quotes)    \\
        2            & Data Date             & データ日付                  & date          & Trading date (YYYYMMDD format, parsed to Date type)                    \\
        3            & Exchange Code         & 取引所コード                 & string        & Stock exchange (e.g., 11 = Tokyo Stock Exchange)                       \\
        4            & Security Type         & 証券種別                   & string        & Market section (e.g., 1 = First Section, 4 = Mothers); the vendor still delivers these pre-2022 segment labels \\
        5            & Session               & 場区分                    & string        & Trading session (1 = Morning 09:00--11:30, 2 = Afternoon 12:30--15:00; --15:30 from 2024-11-05) \\
        6            & Stock Code            & 銘柄コード                  & string        & 4-digit ticker symbol with optional suffix character                   \\
        \midrule
        7            & Execution Time        & 約定時刻                   & time          & Trade execution time (HHMMSS)                                          \\
        8            & Sell Quote Time       & 売り気配時刻                 & time          & Sell-side best quote update time (HHMMSS)                              \\
        9            & Buy Quote Time        & 買い気配時刻                 & time          & Buy-side best quote update time (HHMMSS)                               \\
        10           & Update Time           & 銘柄更新時刻                 & time          & Full-precision update timestamp (HHMMSSNNNNNN)                         \\
        11           & Management Number     & 管理番号                   & string        & Internal sequence number for record ordering                           \\
        \midrule
        12           & Execution Price       & 約定価格                   & float         & Trade price in yen                                                     \\
        13           & Execution Type        & 約定種別                   & string        & Trade classification (Opening, At Buy Quote, Between Quotes, etc.)     \\
        14           & Ayumi Flag            & 歩み値フラグ                 & string        & Tick status (Regular, System Halt, Suspension, Circuit Breaker, etc.)  \\
        15           & Volume                & 売買高                    & int           & Number of shares traded in this execution                              \\
        16           & Volume Flag           & 売買高フラグ                 & string        & Volume status, decoded to ``Final'' (0) or ``Estimated'' (128)         \\
        17           & Close Quote Flag      & 終了時気配フラグ               & string        & End-of-session quote condition indicator                               \\
        \midrule
        18           & Sell Quote 1 Best     & 売り気配１                  & float         & Best ask price (Level 1)                                               \\
        19           & Sell Quote Vol 1      & 売り気配数量１                & int           & Volume at best ask                                                     \\
        20           & Sell Quote Flag 1     & 売り気配フラグ１               & int           & Quote condition flag at best ask                                       \\
        21           & Buy Quote 1 Best      & 買い気配１                  & float         & Best bid price (Level 1)                                               \\
        22           & Buy Quote Vol 1       & 買い気配数量１                & int           & Volume at best bid                                                     \\
        23           & Buy Quote Flag 1      & 買い気配フラグ１               & int           & Quote condition flag at best bid                                       \\
        \midrule
        24           & Sell Limit Quote      & 売り成行気配                 & float         & Sell-side limit order accumulated quote                                \\
        25           & Sell Limit Vol        & 売り成行数量                 & int           & Volume at sell limit quote                                             \\
        26           & Sell Limit Flag       & 売り成行フラグ                & int           & Sell limit quote condition flag                                        \\
        27           & Sell Market Quote     & 売り特別気配                 & float         & Sell-side special quote (tokubetsu kehai)                              \\
        28           & Sell Market Vol       & 売り特別数量                 & int           & Volume at sell special quote                                           \\
        29           & Sell Market Flag      & 売り特別フラグ                & int           & Sell special quote condition flag                                      \\
        \midrule
        30           & Sell Quote 2          & 売り気配２                  & float         & Ask price at depth Level 2                                             \\
        31           & Sell Quote Vol 2      & 売り気配数量２                & int           & Volume at ask Level 2                                                  \\
        32           & Sell Quote Flag 2     & 売り気配フラグ２               & int           & Quote condition flag at ask Level 2                                    \\
        33           & Sell Quote 3          & 売り気配３                  & float         & Ask price at depth Level 3                                             \\
        34           & Sell Quote Vol 3      & 売り気配数量３                & int           & Volume at ask Level 3                                                  \\
        35           & Sell Quote Flag 3     & 売り気配フラグ３               & int           & Quote condition flag at ask Level 3                                    \\
        36           & Sell Quote 4          & 売り気配４                  & float         & Ask price at depth Level 4                                             \\
        37           & Sell Quote Vol 4      & 売り気配数量４                & int           & Volume at ask Level 4                                                  \\
        38           & Sell Quote Flag 4     & 売り気配フラグ４               & int           & Quote condition flag at ask Level 4                                    \\
        39           & Sell Quote 5          & 売り気配５                  & float         & Ask price at depth Level 5                                             \\
        40           & Sell Quote Vol 5      & 売り気配数量５                & int           & Volume at ask Level 5                                                  \\
        41           & Sell Quote Flag 5     & 売り気配フラグ５               & int           & Quote condition flag at ask Level 5                                    \\
        42           & Sell Quote 6          & 売り気配６                  & float         & Ask price at depth Level 6                                             \\
        43           & Sell Quote Vol 6      & 売り気配数量６                & int           & Volume at ask Level 6                                                  \\
        44           & Sell Quote Flag 6     & 売り気配フラグ６               & int           & Quote condition flag at ask Level 6                                    \\
        45           & Sell Quote 7          & 売り気配７                  & float         & Ask price at depth Level 7                                             \\
        46           & Sell Quote Vol 7      & 売り気配数量７                & int           & Volume at ask Level 7                                                  \\
        47           & Sell Quote Flag 7     & 売り気配フラグ７               & int           & Quote condition flag at ask Level 7                                    \\
        48           & Sell Quote 8          & 売り気配８                  & float         & Ask price at depth Level 8                                             \\
        49           & Sell Quote Vol 8      & 売り気配数量８                & int           & Volume at ask Level 8                                                  \\
        50           & Sell Quote Flag 8     & 売り気配フラグ８               & int           & Quote condition flag at ask Level 8                                    \\
        51           & Sell Quote 9          & 売り気配９                  & float         & Ask price at depth Level 9                                             \\
        52           & Sell Quote Vol 9      & 売り気配数量９                & int           & Volume at ask Level 9                                                  \\
        53           & Sell Quote Flag 9     & 売り気配フラグ９               & int           & Quote condition flag at ask Level 9                                    \\
        54           & Sell Quote 10         & 売り気配１０                 & float         & Ask price at depth Level 10                                            \\
        55           & Sell Quote Vol 10     & 売り気配数量１０               & int           & Volume at ask Level 10                                                 \\
        56           & Sell Quote Flag 10    & 売り気配フラグ１０              & int           & Quote condition flag at ask Level 10                                   \\
        \midrule
        57           & Sell Quote OVER       & 売り気配OVER               & float         & Ask overflow price (orders beyond Level 10)                            \\
        58           & Sell Quote Vol OVER   & 売り気配数量OVER             & int           & Volume at ask overflow                                                 \\
        59           & Sell Quote Flag OVER  & 売り気配フラグOVER            & int           & Quote condition flag at ask overflow                                   \\
        \midrule
        60           & Buy Limit Quote       & 買い成行気配                 & float         & Buy-side limit order accumulated quote                                 \\
        61           & Buy Limit Vol         & 買い成行数量                 & int           & Volume at buy limit quote                                              \\
        62           & Buy Limit Flag        & 買い成行フラグ                & int           & Buy limit quote condition flag                                         \\
        63           & Buy Market Quote      & 買い特別気配                 & float         & Buy-side special quote (tokubetsu kehai)                               \\
        64           & Buy Market Vol        & 買い特別数量                 & int           & Volume at buy special quote                                            \\
        65           & Buy Market Flag       & 買い特別フラグ                & int           & Buy special quote condition flag                                       \\
        \midrule
        66           & Buy Quote 2           & 買い気配２                  & float         & Bid price at depth Level 2                                             \\
        67           & Buy Quote Vol 2       & 買い気配数量２                & int           & Volume at bid Level 2                                                  \\
        68           & Buy Quote Flag 2      & 買い気配フラグ２               & int           & Quote condition flag at bid Level 2                                    \\
        69           & Buy Quote 3           & 買い気配３                  & float         & Bid price at depth Level 3                                             \\
        70           & Buy Quote Vol 3       & 買い気配数量３                & int           & Volume at bid Level 3                                                  \\
        71           & Buy Quote Flag 3      & 買い気配フラグ３               & int           & Quote condition flag at bid Level 3                                    \\
        72           & Buy Quote 4           & 買い気配４                  & float         & Bid price at depth Level 4                                             \\
        73           & Buy Quote Vol 4       & 買い気配数量４                & int           & Volume at bid Level 4                                                  \\
        74           & Buy Quote Flag 4      & 買い気配フラグ４               & int           & Quote condition flag at bid Level 4                                    \\
        75           & Buy Quote 5           & 買い気配５                  & float         & Bid price at depth Level 5                                             \\
        76           & Buy Quote Vol 5       & 買い気配数量５                & int           & Volume at bid Level 5                                                  \\
        77           & Buy Quote Flag 5      & 買い気配フラグ５               & int           & Quote condition flag at bid Level 5                                    \\
        78           & Buy Quote 6           & 買い気配６                  & float         & Bid price at depth Level 6                                             \\
        79           & Buy Quote Vol 6       & 買い気配数量６                & int           & Volume at bid Level 6                                                  \\
        80           & Buy Quote Flag 6      & 買い気配フラグ６               & int           & Quote condition flag at bid Level 6                                    \\
        81           & Buy Quote 7           & 買い気配７                  & float         & Bid price at depth Level 7                                             \\
        82           & Buy Quote Vol 7       & 買い気配数量７                & int           & Volume at bid Level 7                                                  \\
        83           & Buy Quote Flag 7      & 買い気配フラグ７               & int           & Quote condition flag at bid Level 7                                    \\
        84           & Buy Quote 8           & 買い気配８                  & float         & Bid price at depth Level 8                                             \\
        85           & Buy Quote Vol 8       & 買い気配数量８                & int           & Volume at bid Level 8                                                  \\
        86           & Buy Quote Flag 8      & 買い気配フラグ８               & int           & Quote condition flag at bid Level 8                                    \\
        87           & Buy Quote 9           & 買い気配９                  & float         & Bid price at depth Level 9                                             \\
        88           & Buy Quote Vol 9       & 買い気配数量９                & int           & Volume at bid Level 9                                                  \\
        89           & Buy Quote Flag 9      & 買い気配フラグ９               & int           & Quote condition flag at bid Level 9                                    \\
        90           & Buy Quote 10          & 買い気配１０                 & float         & Bid price at depth Level 10                                            \\
        91           & Buy Quote Vol 10      & 買い気配数量１０               & int           & Volume at bid Level 10                                                 \\
        92           & Buy Quote Flag 10     & 買い気配フラグ１０              & int           & Quote condition flag at bid Level 10                                   \\
        \midrule
        93           & Buy Quote UNDER       & 買い気配UNDER              & float         & Bid underflow price (orders beyond Level 10)                           \\
        94           & Buy Quote Vol UNDER   & 買い気配数量UNDER            & int           & Volume at bid underflow                                                \\
        95           & Buy Quote Flag UNDER  & 買い気配フラグUNDER           & int           & Quote condition flag at bid underflow                                  \\
    \end{longtable}
\end{landscape}

% ---- TICSS110: full 82-column table ----
\begin{landscape}
    \setlength{\LTcapwidth}{\linewidth}
    \footnotesize
    \begin{longtable}{@{}p{0.7cm} P{4.0cm} P{4.2cm} p{1.0cm} P{9.0cm}@{}}
        \caption{TICSS110 Stock Summary Schema (82 output fields from 83 raw fields; Identification Flag is dropped during processing).} \label{tab:ticss110-full} \\
        \toprule
        \textbf{Col} & \textbf{English Name}             & \textbf{Japanese Name} & \textbf{Type} & \textbf{Description}                                           \\
        \midrule
        \endfirsthead
        \multicolumn{5}{@{}l}{\small\itshape Table~\ref{tab:ticss110-full} continued}                                                                                 \\
        \toprule
        \textbf{Col} & \textbf{English Name}             & \textbf{Japanese Name} & \textbf{Type} & \textbf{Description}                                           \\
        \midrule
        \endhead
        \midrule
        \multicolumn{5}{r@{}}{\small\itshape Continued on next page}                                                                                                  \\
        \endfoot
        \bottomrule
        \endlastfoot
        1            & Record Type                       & レコード種別                 & string        & Record format identifier                                       \\
        2            & Data Date                         & データ日付                  & date          & Trading date                                                   \\
        3            & Exchange Code                     & 取引所コード                 & string        & Stock exchange identifier                                      \\
        4            & Security Type                     & 証券種別                   & string        & Market section code                                            \\
        5            & Stock Code                        & 銘柄コード                  & string        & 4-digit ticker symbol                                          \\
        \midrule
        6            & Trading Unit                      & 単位株数                   & int           & Shares per trading unit                                        \\
        7            & Issued Shares                     & 発行済株式数                 & int           & Total issued share count                                       \\
        \midrule
        8            & Executions $\leq$3 units          & 約定$\leq$3単位            & int           & Trades with $\leq$3 trading units                              \\
        9            & Executions 3$<$x$\leq$6 units     & 約定3$<$x$\leq$6単位       & int           & Trades with 3$<$x$\leq$6 units                                 \\
        10           & Executions 6$<$x$\leq$9 units     & 約定6$<$x$\leq$9単位       & int           & Trades with 6$<$x$\leq$9 units                                 \\
        11           & Executions 9$<$x$\leq$29 units    & 約定9$<$x$\leq$29単位      & int           & Trades with 9$<$x$\leq$29 units                                \\
        12           & Executions 29$<$x$\leq$49 units   & 約定29$<$x$\leq$49単位     & int           & Trades with 29$<$x$\leq$49 units                               \\
        13           & Executions 49$<$x$\leq$99 units   & 約定49$<$x$\leq$99単位     & int           & Trades with 49$<$x$\leq$99 units                               \\
        14           & Executions 99$<$x$\leq$199 units  & 約定99$<$x$\leq$199単位    & int           & Trades with 99$<$x$\leq$199 units                              \\
        15           & Executions 199$<$x$\leq$299 units & 約定199$<$x$\leq$299単位   & int           & Trades with 199$<$x$\leq$299 units                             \\
        \midrule
        16           & AM Opening Price                  & 前場始値                   & float         & Morning session opening price                                  \\
        17           & AM Opening Time                   & 前場始値時刻                 & time          & Morning session opening time                                   \\
        18           & AM Opening Volume                 & 前場始値約定株数               & int           & Shares traded at morning open                                  \\
        19           & AM High Price                     & 前場高値                   & float         & Morning session high price                                     \\
        20           & AM Low Price                      & 前場安値                   & float         & Morning session low price                                      \\
        21           & AM Close Price                    & 前場終値                   & float         & Morning session close price                                    \\
        22           & AM Close Time                     & 前場終値時刻                 & time          & Morning session close time                                     \\
        23           & AM Close Volume                   & 前場終値約定株数               & int           & Shares traded at morning close                                 \\
        24           & AM UpTick Volume                  & 前場値上がり株数               & int           & Shares on uptick trades (AM)                                   \\
        25           & AM UpTick Amount                  & 前場値上がり金額               & float         & Yen amount on uptick trades (AM)                               \\
        26           & AM UpTick Count                   & 前場値上がり回数               & int           & Number of uptick trades (AM)                                   \\
        27           & AM DownTick Volume                & 前場値下がり株数               & int           & Shares on downtick trades (AM)                                 \\
        28           & AM DownTick Amount                & 前場値下がり金額               & float         & Yen amount on downtick trades (AM)                             \\
        29           & AM DownTick Count                 & 前場値下がり回数               & int           & Number of downtick trades (AM)                                 \\
        30           & AM Total Volume                   & 前場約定株数                 & int           & Total shares traded (AM)                                       \\
        31           & AM Total Amount                   & 前場約定金額                 & float         & Total yen traded (AM)                                          \\
        32           & AM Execution Count                & 前場約定回数                 & int           & Total number of executions (AM)                                \\
        33           & AM VWAP                           & 前場VWAP                 & float         & Volume-weighted average price (AM)                             \\
        34           & AM Std Dev                        & 前場標準偏差                 & float         & Price standard deviation (AM)                                  \\
        35           & AM Sell Quote Time                & 前場売り気配時刻               & time          & AM sell quote observation time                                 \\
        36           & AM Buy Quote Time                 & 前場買い気配時刻               & time          & AM buy quote observation time                                  \\
        37           & AM Spread Time                    & 前場スプレッド時刻              & time          & AM spread observation time                                     \\
        38           & AM Avg Sell Quote Vol             & 前場平均売り気配数量             & float         & Average sell quote volume (AM)                                 \\
        39           & AM Avg Buy Quote Vol              & 前場平均買い気配数量             & float         & Average buy quote volume (AM)                                  \\
        40           & AM Avg Spread                     & 前場平均スプレッド              & float         & Average bid--ask spread (AM)                                   \\
        \midrule
        41           & PM Opening Price                  & 後場始値                   & float         & Afternoon session opening price                                \\
        42           & PM Opening Time                   & 後場始値時刻                 & time          & Afternoon session opening time                                 \\
        43           & PM Opening Volume                 & 後場始値約定株数               & int           & Shares traded at afternoon open                                \\
        44           & PM High Price                     & 後場高値                   & float         & Afternoon session high price                                   \\
        45           & PM Low Price                      & 後場安値                   & float         & Afternoon session low price                                    \\
        46           & PM Close Price                    & 後場終値                   & float         & Afternoon session close price                                  \\
        47           & PM Close Time                     & 後場終値時刻                 & time          & Afternoon session close time                                   \\
        48           & PM Close Volume                   & 後場終値約定株数               & int           & Shares traded at afternoon close                               \\
        49           & PM UpTick Volume                  & 後場値上がり株数               & int           & Shares on uptick trades (PM)                                   \\
        50           & PM UpTick Amount                  & 後場値上がり金額               & float         & Yen amount on uptick trades (PM)                               \\
        51           & PM UpTick Count                   & 後場値上がり回数               & int           & Number of uptick trades (PM)                                   \\
        52           & PM DownTick Volume                & 後場値下がり株数               & int           & Shares on downtick trades (PM)                                 \\
        53           & PM DownTick Amount                & 後場値下がり金額               & float         & Yen amount on downtick trades (PM)                             \\
        54           & PM DownTick Count                 & 後場値下がり回数               & int           & Number of downtick trades (PM)                                 \\
        55           & PM Total Volume                   & 後場約定株数                 & int           & Total shares traded (PM)                                       \\
        56           & PM Total Amount                   & 後場約定金額                 & float         & Total yen traded (PM)                                          \\
        57           & PM Execution Count                & 後場約定回数                 & int           & Total number of executions (PM)                                \\
        58           & PM VWAP                           & 後場VWAP                 & float         & Volume-weighted average price (PM)                             \\
        59           & PM Std Dev                        & 後場標準偏差                 & float         & Price standard deviation (PM)                                  \\
        60           & PM Sell Quote Time                & 後場売り気配時刻               & time          & PM sell quote observation time                                 \\
        61           & PM Buy Quote Time                 & 後場買い気配時刻               & time          & PM buy quote observation time                                  \\
        62           & PM Spread Time                    & 後場スプレッド時刻              & time          & PM spread observation time                                     \\
        63           & PM Avg Sell Quote Vol             & 後場平均売り気配数量             & float         & Average sell quote volume (PM)                                 \\
        64           & PM Avg Buy Quote Vol              & 後場平均買い気配数量             & float         & Average buy quote volume (PM)                                  \\
        65           & PM Avg Spread                     & 後場平均スプレッド              & float         & Average bid--ask spread (PM)                                   \\
        \midrule
        66           & Daily VWAP                        & 全日VWAP                 & float         & Full-day volume-weighted average price                         \\
        67           & Daily Std Dev                     & 全日標準偏差                 & float         & Full-day price standard deviation                              \\
        68           & Daily Weighted Avg Sell Quote     & 全日加重平均売り気配             & float         & Full-day weighted average sell quote                           \\
        69           & Daily Weighted Avg Buy Quote      & 全日加重平均買い気配             & float         & Full-day weighted average buy quote                            \\
        70           & Daily Avg Spread                  & 全日平均スプレッド              & float         & Full-day average bid--ask spread                               \\
        \midrule
        71           & AM Sell Quote Execution Vol       & 前場売り気配約定株数             & int           & AM shares executed at sell quote                               \\
        72           & AM Sell Quote Execution Amt       & 前場売り気配約定金額             & float         & AM yen amount executed at sell quote                           \\
        73           & AM Sell Quote Execution Cnt       & 前場売り気配約定回数             & int           & AM execution count at sell quote                               \\
        74           & AM Buy Quote Execution Vol        & 前場買い気配約定株数             & int           & AM shares executed at buy quote                                \\
        75           & AM Buy Quote Execution Amt        & 前場買い気配約定金額             & float         & AM yen amount executed at buy quote                            \\
        76           & AM Buy Quote Execution Cnt        & 前場買い気配約定回数             & int           & AM execution count at buy quote                                \\
        77           & PM Sell Quote Execution Vol       & 後場売り気配約定株数             & int           & PM shares executed at sell quote                               \\
        78           & PM Sell Quote Execution Amt       & 後場売り気配約定金額             & float         & PM yen amount executed at sell quote                           \\
        79           & PM Sell Quote Execution Cnt       & 後場売り気配約定回数             & int           & PM execution count at sell quote                               \\
        80           & PM Buy Quote Execution Vol        & 後場買い気配約定株数             & int           & PM shares executed at buy quote                                \\
        81           & PM Buy Quote Execution Amt        & 後場買い気配約定金額             & float         & PM yen amount executed at buy quote                            \\
        82           & PM Buy Quote Execution Cnt        & 後場買い気配約定回数             & int           & PM execution count at buy quote                                \\
    \end{longtable}
\end{landscape}

% ---- TICIT110: full 10-column table ----
\begin{table}[p]
    \centering
    \small
    \caption{TICIT110 Index Tick Data Schema (10 output fields selected from 23 raw fields; 13 reserved fields are dropped).} \label{tab:ticit110-full}
    \begin{tabular}{@{}c P{3cm} P{2.9cm} c P{5.1cm}@{}}
        \toprule
        \textbf{Col} & \textbf{English Name} & \textbf{Japanese Name} & \textbf{Type} & \textbf{Description}                                   \\
        \midrule
        1            & Record Type           & レコード種別                 & string        & Record format identifier                               \\
        2            & Data Date             & データ日付                  & date          & Trading date                                           \\
        3            & Exchange Code         & 取引所コード                 & string        & Exchange identifier                                    \\
        4            & Security Type         & 証券種別                   & string        & Security type code (10 = Cash Index)                   \\
        5            & Session               & 場区分                    & string        & Trading session (1 = AM, 2 = PM)                       \\
        6            & Index Code            & 指数コード                  & string        & Index identifier (e.g., 101 = Nikkei 225, 113 = TOPIX) \\
        7            & Execution Time        & 約定時刻                   & time          & Index value calculation time (HHMMSS)                  \\
        8            & Index Value           & 指数値                    & float         & Index level (scaled $\times 0.01$ from raw integer)    \\
        9            & Execution Type        & 約定種別                   & string        & Record type (1 = Opening, 2 = Post-Closing, 0 = Other) \\
        10           & Ayumi Flag            & 歩み値フラグ                 & string        & Tick status (0 = Regular, 128 = Closing)               \\
        \bottomrule
    \end{tabular}
\end{table}

% ---- TICIS110: full 17-column table ----
\begin{table}[p]
    \centering
    \small
    \caption{TICIS110 Index Summary Schema (17 output fields selected from 83 raw fields in the 2017-onward era; the 2016 era is a fixed-width hybrid with the same output).} \label{tab:ticis110-full}
    \begin{tabular}{@{}c P{3cm} P{2.9cm} c P{5.1cm}@{}}
        \toprule
        \textbf{Col} & \textbf{English Name} & \textbf{Japanese Name} & \textbf{Type} & \textbf{Description}                                   \\
        \midrule
        1            & Record Type           & レコード種別                 & string        & Record format identifier                               \\
        2            & Data Date             & データ日付                  & date          & Trading date                                           \\
        3            & Exchange Code         & 取引所コード                 & string        & Exchange identifier                                    \\
        4            & Security Type         & 証券種別                   & string        & Security type code (10 = Cash Index)                   \\
        5            & Index Code            & 指数コード                  & string        & Index identifier (e.g., 101 = Nikkei 225, 113 = TOPIX) \\
        \midrule
        6            & AM Opening Price      & 前場始値                   & float         & Morning session opening index level                    \\
        7            & AM Opening Time       & 前場始値時刻                 & time          & Morning session opening time (HHMMSS)                  \\
        8            & AM High Price         & 前場高値                   & float         & Morning session high                                   \\
        9            & AM Low Price          & 前場安値                   & float         & Morning session low                                    \\
        10           & AM Close Price        & 前場終値                   & float         & Morning session close                                  \\
        11           & AM Close Time         & 前場終値時刻                 & time          & Morning session close time (HHMMSS)                    \\
        \midrule
        12           & PM Opening Price      & 後場始値                   & float         & Afternoon session opening index level                  \\
        13           & PM Opening Time       & 後場始値時刻                 & time          & Afternoon session opening time (HHMMSS)                \\
        14           & PM High Price         & 後場高値                   & float         & Afternoon session high                                 \\
        15           & PM Low Price          & 後場安値                   & float         & Afternoon session low                                  \\
        16           & PM Close Price        & 後場終値                   & float         & Afternoon session close                                \\
        17           & PM Close Time         & 後場終値時刻                 & time          & Afternoon session close time (HHMMSS)                  \\
        \bottomrule
    \end{tabular}
\end{table}